\documentclass[twocolumn,prx,amsmath,amssymb,superscriptaddress]{revtex4-2}
\usepackage{graphicx}
\graphicspath{
    {figures/}
}
\usepackage{subcaption}
\usepackage{ragged2e}
\DeclareCaptionJustification{justified}{\justifying}
\usepackage{hyperref}
\usepackage{cleveref}
\usepackage{physics}

\begin{document}
    \title{Manifestation of the coupling phase in microwave cavity magnonics}

    \author{Alan Gardin}
    \email{alan.gardin@adelaide.edu.au}
    \affiliation{%
    	School of Physics, The University of Adelaide, Adelaide SA 5005, Australia
    }%
	\affiliation{%
		IMT Atlantique, Technopole Brest-Iroise, CS 83818, 29238 Brest Cedex 3, France
	}%
	\affiliation{%
		Lab-STICC (UMR 6285), CNRS, Technopole Brest-Iroise, CS 83818, 29238 Brest Cedex 3, France
	}%
	\author{Jeremy Bourhill}
	\affiliation{%
		IMT Atlantique, Technopole Brest-Iroise, CS 83818, 29238 Brest Cedex 3, France
	}%
	\affiliation{%
		ARC Centre of Excellence for Engineered Quantum Systems and ARC Centre of Excellence for Dark Matter Particle Physics, \\
		Department of Physics, University of Western Australia,	35 Stirling Highway, Crawley, Western Australia 6009, Australia
	}%
	\author{Vincent Vlaminck}
	\affiliation{%
		IMT Atlantique, Technopole Brest-Iroise, CS 83818, 29238 Brest Cedex 3, France
	}%
	\affiliation{%
		Lab-STICC (UMR 6285), CNRS, Technopole Brest-Iroise, CS 83818, 29238 Brest Cedex 3, France
	}%
    \author{Christian Person}
    \affiliation{%
    	IMT Atlantique, Technopole Brest-Iroise, CS 83818, 29238 Brest Cedex 3, France
    }%
    \affiliation{%
    	Lab-STICC (UMR 6285), CNRS, Technopole Brest-Iroise, CS 83818, 29238 Brest Cedex 3, France
    }%
    \author{Christophe Fumeaux}
    \affiliation{%
    	School of Electrical and Electronic Engineering, The University of Adelaide, Adelaide SA 5005, Australia
    }%
    \author{Vincent Castel}
    \affiliation{%
    	IMT Atlantique, Technopole Brest-Iroise, CS 83818, 29238 Brest Cedex 3, France
    }%
    \affiliation{%
        Lab-STICC (UMR 6285), CNRS, Technopole Brest-Iroise, CS 83818, 29238 Brest Cedex 3, France
    }%
    \author{Giuseppe C. Tettamanzi}
    \affiliation{%
    	School of Physics, The University of Adelaide, Adelaide SA 5005, Australia
    }%
	\affiliation{%
		School of Chemical Engineering and Advanced Materials, The University of Adelaide, Adelaide SA 5005, Australia
	}%

    \begin{abstract}
    	The interaction between microwave photons and magnons is well understood and originates from the Zeeman coupling between spins and a magnetic field. Interestingly, the magnon/photon interaction is accompanied by a phase factor which can usually be neglected. However, under the rotating wave approximation, if two magnon modes simultaneously couple with two cavity resonances, this phase cannot be ignored as it changes the physics of the system. We consider two such systems, each differing by the sign of one of the magnon/photon coupling strengths. This simple difference, originating from the various coupling phases in the system, is shown to preserve, or destroy, two potential applications of hybrid photon/magnon systems, namely dark mode memories and cavity-mediated coupling. The observable consequences of the coupling phase in this system is akin to the manifestation of a discrete Pancharatnam–Berry phase, which may be useful for quantum information processing.
    \end{abstract}
    \maketitle

    \section{Introduction}
    Magnons are quasi-particles associated with the collective excitation of spins in magnetic materials. Magnons in ferrimagnetic insulators such as Yttrium-Iron-Garnet (YIG) are promising for information transduction, due to their ability to couple to a plethora of systems, such as mechanical resonators or optical and microwave photonic modes \cite{2016Zhang,2019LachanceQuirion}. The emerging field of cavity magnonics focuses on the photon/magnon interaction confined within cavities at microwave or optical frequencies \cite{2018HarderHu,2021Harder,2021Rameshti}. In the microwave domain, strong and ultra-strong coherent coupling have been demonstrated \cite{2014Zhang,2016Bourhill,2014Goryachev,2021Golovchanskiy,2021Golovchanskiya}, with the latter even reachable at room temperature \cite{2022Bourcin}.
    
    Owing to this flexibility, indirect coupling of two non-interacting systems by an auxiliary mode has been investigated in \cite{2016Hyde,2021Crescini,2022An,2016Lambert,2018Rameshti,2019Grigoryan,2019Xu}. For instance, two macroscopically distant magnetic samples in a cavity can indirectly couple by using the cavity photons as a bridge \cite{2016Lambert,2018Rameshti,2019Grigoryan,2019Xu,2021Zhan,2022Nair,2015Zhang}. Notably, in this configuration some eigenmodes are ``dark'', in the sense that they do not lead to an experimental signature when probing the system. This dark mode physics can be used to create dark mode gradient memories as experimentally demonstrated in \citeauthor{2015Zhang}. Another example of cavity-mediated coupling is the coupling of a magnon with a superconducting qubit, again mediated by a common cavity mode \cite{2015Tabuchi,2016Tabuchi}. This versatile configuration motivated studies of quantum magnonics \cite{2022Yuan}, with proposals for single-magnon sources \cite{2019Liu,2020Xie}, multi-magnon blockade \cite{2021Wu}, single-shot detection of a single magnon \cite{2020LachanceQuirion} and quantum sensing of magnons \cite{2020Wolski}.
    
    These applications rely on the coupling between microwave photons and magnons, which physically originates from the Zeeman coupling between the spins in the ferrimagnetic material and the cavity's RF magnetic field \cite{2019FlowerGoryachevBourhillTobar}. Adopting a quantum formalism to describe this coupling allows for the precise computation of the coupling strength based on microscopic parameters and the cavity geometry \cite{2020BourhillCastel}. Interestingly, the magnon/photon coupling term resulting from the Zeeman coupling term is accompanied by a phase factor. To the authors' knowledge, this phase has never been discussed explicitly in the literature, probably due to its inconspicuous nature in the systems considered so far. 
    
    The main contribution of this paper is to highlight how the coupling phases can become relevant in systems composed of several magnon and cavity modes. To that effect, we begin by explaining the origin of the coupling phase between a microwave photon and a magnon in \cref{sec:coupling-phase}. In \cref{sec:model}, we introduce a hybrid system composed of two YIG spheres, each coupling to two magnetic eigenmodes of a microwave cavity. We show that the various magnon/photon coupling phases in such a system lead to an observable quantity $\theta$ parametrising the Hamiltonian. To illustrate the impact of the physical phase $\theta$, we study the physics of such a system in \cref{sec:influence-theta}, and find that both cavity-mediated coupling and dark mode physics are $\theta$-dependent.

	\section{\label{sec:coupling-phase}Coupling phase between a microwave photon and a magnon}
	\subsection{Free Hamiltonian}
	In this section, we consider the interaction between one YIG sphere and one magnetic eigenmode of a microwave cavity. The geometry of the cavity fixes the resonance frequency $\omega_c/2\pi$ of the magnetic mode $\vb{H}$. These two quantities can be obtained by solving Maxwell's equations, for instance using an electromagnetic finite-element modelling software such as COMSOL Multiphysics\textsuperscript{\textregistered}. 
	
	The lowest-order magnetostatic mode in the YIG sphere (the Kittel mode, $\vb{k}=0$) corresponds to a collective precession of the spins, which can be described using a macroscopic spin \cite{2010SoykalPRB,2010SoykalPRL}. Whilst higher-order standing wave modes (for $\vb{k}\neq 0$) exist \cite{2009Stancil}, we focus on the fundamental uniform mode in this work. For a spherical YIG sample, the ferromagnetic resonance (FMR) frequency $\omega_m/2\pi$ can be tuned by an applied static magnetic field $\vb{H}_0 = H_0 \vu{z}$ as $\omega_m = \gamma \abs{\vb{H}_0}$, with $\gamma$ the gyromagnetic ratio.
	
	After quantisation of the electromagnetic field, the cavity mode takes the form of a quantised harmonic oscillator $\hbar \omega_c c^\dagger c$, described using bosonic annihilation and creation operators $c$ and $c^\dagger$ respectively. Provided the number of magnons is negligible compared to the number of spins in the YIG, we can describe the magnon by a bosonic annihilation operator $m$ after a Holstein-Primakoff transformation \cite{1940Holstein}, leading to a harmonic oscillator for the magnon mode \cite{2009Stancil}. 
    The resulting free Hamiltonian (i.e. without interactions) is the sum of two quantised harmonic oscillators
  	\begin{equation}
  		\label{eq:free-hamiltonian-two-modes}
  		H_\text{free} =
  		\hbar \omega_{c} c^\dagger c
  		+ \hbar \omega_{m} m^\dagger m,
  	\end{equation}
  	describing the cavity and magnon mode respectively. Note that these operators commute.

    \subsection{Interaction term}
    The magnetic mode in the cavity, $\vb{H}$, couples to the macrospin of the YIG sphere by a Zeeman interaction. 
    Assuming the cavity mode $\vb{H}$ has no $z$ dependence (for instance by considering a re-entrant cavity \cite{2019FlowerGoryachevBourhillTobar,2020BourhillCastel,2016Carvalho,Tobar}), the interaction term reads (details in \cref{app:derivation-form-factor})
    \begin{equation}
		\label{eq:interaction-complex-simple-no-RWA}
		H_I
		=
		\hbar ge^{i\varphi} \qty(c + c^\dagger) m^\dagger + h.c,
	\end{equation}
	where the coupling strength $g/2\pi >0 $ and coupling phase $\varphi > 0$ are real numbers, and $h.c$ denotes the omitted hermitian conjugate terms. The coupling strength has the well-known expression \cite{2019FlowerGoryachevBourhillTobar,2020BourhillCastel}
	\begin{equation}
		\label{eq:coupling-strength-single}
		\frac{g}{2\pi} = \eta \sqrt{\omega_{c}} \frac{\gamma}{4\pi} \sqrt{\frac{\mu}{g_L \mu_B} \mu_0 \hbar n_s}
	\end{equation}
	where $\mu_B$ is the Bohr magneton, $\mu = 5 \mu_B$ the magnetic moment of YIG, $g_L=2$ the Landé $g$-factor, $\mu_0$ the magnetic permeability of vacuum, $n_s = 4.22 \times 10^{27} \text{ m}^{-3}$ the spin density of YIG, and the so-called form-factor
	\begin{equation}
		\label{eq:form-factor-single}
		\eta = 
		\sqrt{
			\frac{
				\qty(\int_{V_{m}} \vb{H} \cdot \vu{x} \dd[3]{r})^2 + \qty(\int_{V_{m}} \vb{H} \cdot \vu{y} \dd[3]{r})^2
			}{V_{m} \int_{V_c} \abs{\vb{H}}^2 \dd[3]{r}}
		},
	\end{equation}
	with $V_c$ the volume of the cavity and $V_m$ the volume of the magnetic sample. 
	On the other hand, the coupling phase $\varphi$ reads
	\begin{equation}
		\label{eq:coupling-phase}
		\varphi = \arg \widetilde{H},
	\end{equation}
	where
	\begin{equation}
		\label{eq:definition-H-tilde}
		\widetilde{H} = \int_{V_{m}} \vb{H} \cdot \vu{x} \dd[3]{r} + i \int_{V_{m}} \vb{H} \cdot \vu{y} \dd[3]{r}.
	\end{equation}
	From the definition of $\widetilde{H}$, we see that the coupling phase depends on the orientation of the cavity's magnetic mode $\vb{H}$ traversing the magnetic sample. 
	%\unsure{Note that we need to choose a reference phase to fix the value of $\varphi$.} This choice is phase is arbitrary and suggests a $U(1)$ gauge degree of freedom local to any magnon mode coupling to a microwave photon.
	
	For moderate values of the coupling strength $g/2\pi$, we can perform the rotating wave approximation (RWA) and neglect the counter-rotating terms $cm + c^\dagger m^\dagger$ in \cref{eq:interaction-complex-simple-no-RWA}, so that the interaction simplifies to
	\begin{equation}
		\label{eq:interaction-complex-simple}
		H_I
		=
		\hbar ge^{i\varphi} c m^\dagger + h.c,
	\end{equation}
	which is valid in the strong coupling regime ($g/\omega_c < 0.1$), but not in the ultrastrong coupling regime ($g/\omega_c \geqslant 0.1$) \cite{2022Bourcin}. We refer the reader to \cite{2019FornDiaz,2019FriskKockum,2020LeBoite} for details about the ultrastrong coupling regime and the applicability of the RWA. In the remainder of this paper, we will always apply the RWA, and consider \cref{eq:interaction-complex-simple} instead of \cref{eq:interaction-complex-simple-no-RWA}.
	
	To conclude this section, we would like to mention that the coupling phase $\varphi$ is fundamentally different from the phase factor used to model dissipative couplings \cite{2020Wang}. Indeed, the magnon/photon dissipative coupling can be modelled as
	\begin{equation}
		H_{I,\text{dissipative}} = \hbar J(cm^\dagger + e^{i\phi} c^\dagger m),
	\end{equation}
	where $J$ and $\phi$ are constants. We notice that this term is not hermitian, contrary to \cref{eq:interaction-complex-simple}. As a result, a dissipatively coupled system can have complex eigenvalues, which are at the origin of energy level attraction in the spectrum. Conversely, a hermitian system guarantees real eigenvalues, and the coupling manifests as the standard level repulsion instead, or in other words an anti-crossing in the spectrum.

    \section{\label{sec:model}Example with a physical phase}
    The previous section highlights the presence of a coupling phase $\varphi$ for each magnon/photon coupling term. In this section, we consider a system under the rotating wave approximation composed of two identical YIG spheres, each coupling to two magnetic eigenmodes of the cavity. The free Hamiltonian of the previous section is naturally generalised to
    \onecolumngrid
    
    \begin{center}
    	\begin{figure}[h]
    		\includegraphics[width=\linewidth]{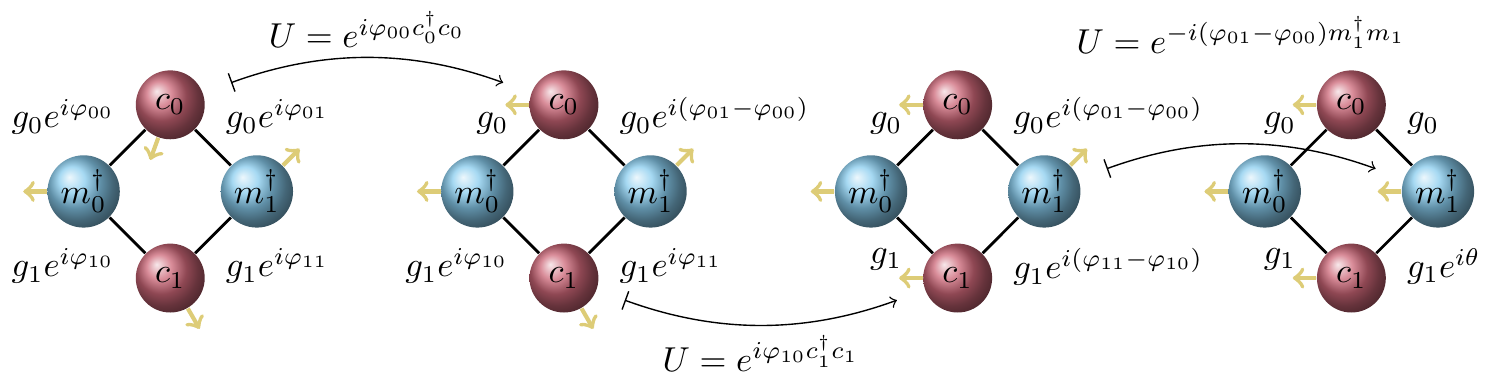}
    		\caption{Illustration of the appearance of a physical phase $\theta$ when two magnon modes $m_0,m_1$ (blue circles) both couple to two cavity modes $c_0,c_1$ (red circles). The black lines between each circle represent a coupling described by \cref{eq:interaction-complex-simple} (the hermitian conjugate term is omitted for readability). The coupling phase measures the difference in orientation of the yellow arrows coming out of each circle. By moving to an appropriate reference frame using a unitary transformation, we can align the yellow arrows of two interacting modes, hence removing the coupling phase. We successively (from left to right) rotate the cavity mode $c_0$, the cavity mode $c_1$, and finally the magnon mode $m_1$ to align the yellow arrows with that of $m_0$. After successive rotation of the operators, all yellow arrows point in the same direction, and yet a physical phase $\theta = \varphi_{11}-\varphi_{01} - \qty(\varphi_{10}-\varphi_{00})$ remains. Note that the free Hamiltonian is unaffected by the successive rotation of the modes, see \cref{app:rotating-frame}.}
    		\label{fig:physical-phase}
    	\end{figure}
    \end{center}
    \twocolumngrid\noindent
    \begin{equation}
    	\label{eq:modelX-free-hamiltonian}
    	H_\text{free} =
    	\hbar \omega_{c,0} c_0^\dagger c_0
    	+ \hbar \omega_{c,1} c_1^\dagger c_1
    	+ \hbar \omega_{m,0} m_0^\dagger m_0
    	+ \hbar \omega_{m,1} m_1^\dagger m_1,
    \end{equation}
	where the $c_k$ describe the two magnetic eigenmodes of the cavity, and $m_k$ the two magnon modes associated with each YIG sphere. Again, the operators $c_k$ and $m_k$ all commute with each other, as they describe independent harmonic oscillators. We introduce the notations $\omega_o$ and $\delta_o$ for each operator $o \in \qty{c,m}$ as
	\begin{equation}
		\omega_o = \frac{\omega_{o,0}+\omega_{o,1}}{2}, \quad
		\delta_o = \frac{\omega_{o,1}-\omega_{o,0}}{2},
	\end{equation}
	so that
	\begin{equation}
		\omega_{o, 0} = \omega_o - \delta_o, \quad
		\omega_{o, 1} = \omega_o + \delta_o.
	\end{equation}
	The microwave cavity design determines the two resonances of the cavity modes, and hence the detuning $\delta_c$ between them and the average cavity resonance $\omega_c$. For the magnons, the applied magnetic field $\vb{H}_0$ tunes the average magnon frequency ${\omega}_m = \gamma \abs{\vb{H_0}}$, whilst the magnon detuning $\delta_m$ can be created by a permanent magnet or a coil placed near the YIG spheres (such a setup was used in \cite{2015Zhang} for instance).
    
    Assuming that both magnon modes couple with the same coupling strength $g_k$ to the cavity mode $c_k$, the interaction Hamiltonian under the RWA is generalised to
    \begin{equation}
    	\label{eq:interaction-hamiltonian-general}
    	\begin{aligned}
    		H_I
    		&=
    		\hbar g_0e^{i\varphi_{00}} c_0 m_0^\dagger + \hbar g_0e^{i\varphi_{01}} c_0 m_1^\dagger +\\
    		&\quad
    		\hbar g_1e^{i\varphi_{10}} c_1 m_0^\dagger + \hbar g_1e^{i\varphi_{11}} c_1 m_1^\dagger
    		+ h.c.
    	\end{aligned}
    \end{equation}
	This interaction Hamiltonian can be simplified by focusing all the coupling phases into one term (see \cref{fig:physical-phase}) using unitary transformations (we recall some elementary properties of unitary transformations in \cref{app:rotating-frame}). 
	The transformed interaction Hamiltonian is
    \begin{equation}
    	\label{eq:interaction-hamiltonian-general-single-phase}
    	\begin{aligned}
    		H_{I,\theta}
    		&=
    		\hbar g_0 c_0 m_0^\dagger + \hbar g_0 c_0 m_1^\dagger \\
    		&\quad
    		+\hbar g_1 c_1 m_0^\dagger + \hbar g_1e^{i\theta} c_1 m_1^\dagger
    		+ h.c,
    	\end{aligned}
    \end{equation}
	with $\theta = \varphi_{11}-\varphi_{01} - \qty(\varphi_{10}-\varphi_{00})$. The ``physical phase'' $\theta$ is reminiscent of a discrete Pancharatnam–Berry phase, a gauge-invariant geometrical phase factor originating from the existence of a local $U(1)$ gauge degree of freedom for each state involved in a loop in some parameter space \cite{2000Resta}. This $U(1)$ gauge is nothing but a choice of phase, which in our system is the coupling phase located on the black lines of \cref{fig:physical-phase}. The various coupling phases are linked to each other through the circles, forming a loop -- leading to an observable physical phase $\theta$.
	
	%As explained in the previous section, the individual coupling phases $\varphi_{ij}$ appearing in \cref{eq:interaction-hamiltonian-general} depend on the choice of a reference phase local to the magnon mode $j$. However, the successive dephasings $\varphi_{1j}-\varphi_{0j}$ eliminate this local gauge degree of freedom, so that the quantity $\theta$ becomes gauge-invariant and hence observable. 
    %Note that the local choices of reference phase for the magnon modes do not affect the value of $\theta$, so that the $U(1)$ degree of freedom local to each magnon remains observable. This phenomenon is akin to a discrete version of the Pancharatnam–Berry phase, a gauge-invariant geometrical phase factor originating from the adiabatic evolution of a state over a loop in some parameter space \cite{2000Resta}. 
    
	To conclude this section, we note that most systems studied in the literature consists of single magnon/photon systems, or several magnon modes coupling to a single cavity mode. In particular, setups comprised of two magnon modes coupling to one cavity mode have attracted interest for indirect coupling and dark mode physics \cite{2015Zhang,2016Lambert,2018Rameshti,2019Grigoryan,2021Zhan,2022Nair}, as well as the generation of entangled states \cite{2019LiZhu,2019ZhangScullyAgarwal,2020Nair}. In these cases, the coupling phases do not lead to a physical phase, as we show in \cref{app:relevance-phase}. This is in stark contrast with the system proposed here.

	\section{\label{sec:influence-theta}Influence of the physical phase}
	We now highlight the $\theta$-dependent physics of the system described in the previous section, i.e. 
	\begin{equation}
		\label{eq:H-theta}
		H_\theta = H_\text{free} + H_{I,\theta},
	\end{equation}
	with $H_\text{free}$ and $H_{I,\theta}$ given by \cref{eq:modelX-free-hamiltonian,eq:interaction-hamiltonian-general-single-phase}. While $\theta$ could in principle take any value in $\qty[0, 2\pi]$, we will focus our analysis on $\theta=0$ and $\theta=\pi$, 
	differing only in the sign of the coupling strength of the last term.
	%	\begin{align}
		%		\label{eq:modelA-interaction}
		%		H_{I,\theta=0} &= 
		%		\hbar g_0 c_0 m_0^\dagger + \hbar g_0 c_0 m_1^\dagger \nonumber\\
		%		&\quad
		%		+ \hbar g_1 c_1 m_0^\dagger + \hbar g_1 c_1 m_1^\dagger + h.c,\\
		%		\label{eq:modelB-interaction}
		%		H_{I,\theta=\pi} &= 
		%		\hbar g_0 c_0 m_0^\dagger + \hbar g_0 c_0 m_1^\dagger \nonumber\\
		%		&\quad
		%		+ \hbar g_1 c_1 m_0^\dagger - \hbar g_1 c_1 m_1^\dagger + h.c,
		%	\end{align}

	\subsection{\label{subsec:diagonalisation-method}Diagonalisation method}
	The coupling between the photonic and matter degrees of freedom described by \cref{eq:interaction-hamiltonian-general-single-phase} leads to the appearance of quasiparticles known as polaritons. Polaritons are hybridised light/matter states and, in the context of cavity magnonics, they are known as cavity magnon-polaritons (CMP) \cite{2016Harder,2018HarderHu}. 
	Diagonalisation of the system allows one to find the frequencies of the polaritons, as well as their composition in terms of the cavity and magnon mode operators $c_k, m_k$. All the modes of the system being bosonic, diagonalisation can be achieved by writing $H_\theta = \hbar \vb*{v}^\dagger A \vb*{v}$ with $\vb*{v} = \mqty(c_0, c_1, m_0, m_1)^t$ ($^t$ indicates matrix transposition) and
	\begin{equation}
		\label{eq:matrix-M-def}
		A =\mqty(
		\omega_{c, 0} & 0 & g_{0} & g_{0} \\
		0 & \omega_{c, 1} & g_{1} & g_{1}e^{-i\theta} \\
		g_{0} & g_{1} & \omega_{m,0} & 0\\
		g_{0} & g_{1}e^{i\theta} & 0 & \omega_{m,1}
		).
	\end{equation}
	After diagonalisation of $A$, the eigenvectors give the polaritonic operators $p_\mu$, while the eigenvalues give the polaritonic frequencies $\omega_\mu/2\pi$. The polaritons can be expressed in terms of the  initial operators as
	\begin{equation}
		\label{eq:polariton-expansion}
		p_\mu = u_{\mu 0} c_0 + u_{\mu 1} c_1 + v_{\mu 2} m_0 + v_{\mu 3} m_1,
	\end{equation}
	where the coefficients $u$ ($v$) are the first (last) two components of the eigenvectors of $A$. 
	
	We note that in general, diagonalising $A$ amounts to solving a polynomial equation of degree four, namely
	\begin{equation}
		\label{eq:eigenvalue-eq}
		\det(A - \omega_\mu I_4)
		= 0
	\end{equation}
	where $I_4$ is the $4 \times 4$ identity matrix. Whilst analytical solutions exist, they are rather unpleasant and do not clarify the physics.
	
%	Finally, note that the procedure described here can be used by choosing another basis, by adapting the definition of $\vb*{v}$. Using an alternative basis will result in identical eigenvalues, but different eigenvectors.

	\begin{figure}[t]
		\centering
		\includegraphics[width=\linewidth]{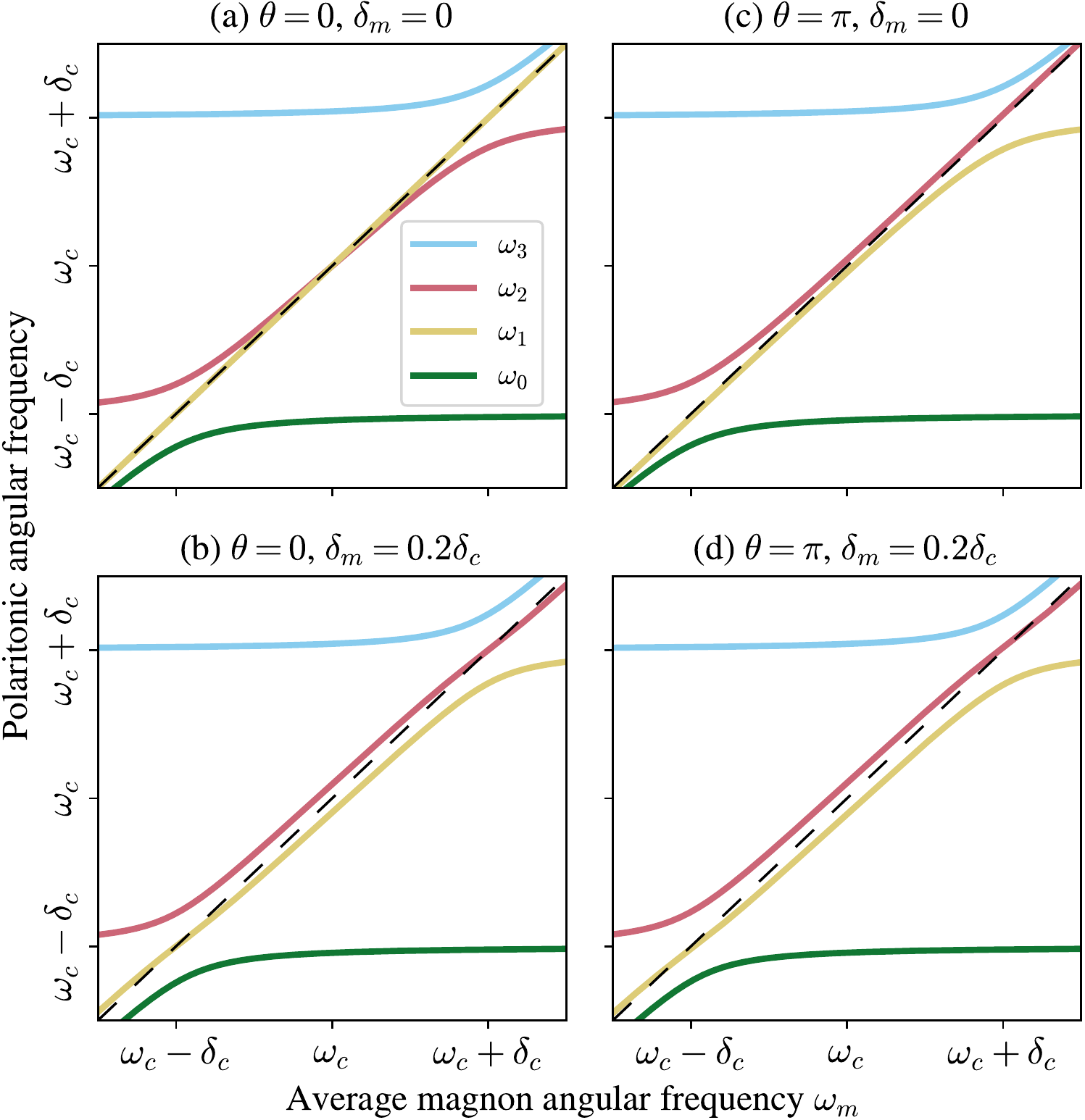}
		\caption{Numerical spectra of $H_\theta$ for $\theta=0$ and $\theta=\pi$ as a function of the average magnon frequency $\omega_m/2\pi$, for two values of the magnon detuning $\delta_m$. Parameters are $\omega_{c,0}=4$ GHz, $\omega_{c, 1}=6$ GHz and $g_0 = g_1 = 0.03 \omega_{c}$. The dashed line corresponds to $\omega = \omega_m$. The legend in (a) is common to (b), (c) and (d).}
		\label{fig:spectra}
		\phantomsubcaption\label{fig:spectra:modelA}
		\phantomsubcaption\label{fig:spectra:modelA-detuned}
		\phantomsubcaption\label{fig:spectra:modelB}
		\phantomsubcaption\label{fig:spectra:modelB-detuned}
	\end{figure}

	\subsection{\label{subsec:spectrum}Spectrum and cavity-mediated coupling}
	\paragraph{Spectral features.}
	We numerically diagonalised \cref{eq:matrix-M-def} for $\theta=0$ and $\theta=\pi$ in \cref{fig:spectra}. The coherent magnon/photon couplings manifest as two anti-crossings, each located near the resonance of the associated cavity mode (i.e. at $\omega_c \pm \delta_c$). We observe that the value of the physical phase $\theta$ seems to have an impact only when $\delta_m=0$, since as $\delta_m$ increases the cases $\theta=0$ and $\theta=\pi$ become less distinguishable. 
	
	Interestingly, it appears that when $\delta_m=0$ and $\omega_m = \omega_c$, the eigenvalues $\omega_1$ and $\omega_2$ cross for $\theta=0$ (\cref{fig:spectra:modelA}), but do not for $\theta=\pi$ (\cref{fig:spectra:modelB}). Repulsion between energy levels is usually the signature of a coupling phenomenon, while level crossing is indicative of the absence of coupling. This suggests that at $\omega_m=\omega_c$, the eigenmodes associated with $\omega_1$ and $\omega_2$ couple for $\theta=\pi$, but do not for $\theta=0$.
	
	\paragraph{Analysis in the dispersive regime.}
	To clarify this spectral feature, we consider the dispersive regime, where both magnons are assumed to be far detuned from both cavity modes. Defining the detuning $\Delta_{kk'} =\omega_{m,k'}-\omega_{c,k}$ between the cavity mode $c_k$ and the magnon mode $m_{k'}$, the dispersive limit corresponds to $\abs{\Delta_{kk'}} \gg g_{k}$ for all $k,k' \in \qty{0,1}$. Under this approximation, we can employ a Schrieffer-Wolff transformation to perform first-order perturbation theory in $\abs{\frac{g_k}{\Delta_{kk'}}} \ll 1$ (see \cref{app:schrieffer-wolff}). We find that the magnon modes being significantly detuned from the cavity modes, the photon/magnon couplings are negligible, and the cavity and magnon modes decouple from each other (see \cref{eq:disperive-hamiltonian}). However, virtual photons still mediate a magnon/magnon interaction, which for $\omega_m=\omega_c$, is described by the effective Hamiltonian (again, see \cref{app:schrieffer-wolff} for details)
	\begin{equation}
		\label{eq:dispersive-hamiltonian-magnons}
		\begin{aligned}
			H'_\text{magnons} 
			&= 
			\hbar \omega'_{m,0} m_0^\dagger m_0
			+ \hbar \omega'_{m,1} m_1^\dagger m_1\\
			&\quad
			+ \hbar \qty(G_\theta m_0 m_1^\dagger + h.c),
		\end{aligned}
	\end{equation}
	where the frequencies of the magnons are shifted due to the cavity modes as
	\begin{equation}
		\label{eq:wm-pulled}
		\omega'_{m,k} = \omega_c + \frac{g_0^2}{\Delta_{0k}} + \frac{g_1^2}{\Delta_{1k}}
	\end{equation}
	and the indirect magnon-magnon coupling $G_\theta/2\pi$ is 
	\begin{equation}
		\label{eq:magnon-magnon-coupling}
		G_\theta = \frac{\delta_c}{\delta_c^2 - \delta_m^2} \qty(g_0^2 - e^{i\theta} g_1^2).
	\end{equation}
	
	The magnitude of $G_\theta$ characterises the strength of the magnon-magnon coupling, and predicts level repulsion of the eigenmodes of \cref{eq:dispersive-hamiltonian-magnons}. However, for $g_0=g_1$ and $\theta = 0$, the magnons do not interact since $G_\theta$ vanishes, leading to crossing of the energy levels. 
	
\begin{figure}[t]
		\centering
		\includegraphics[width=\columnwidth]{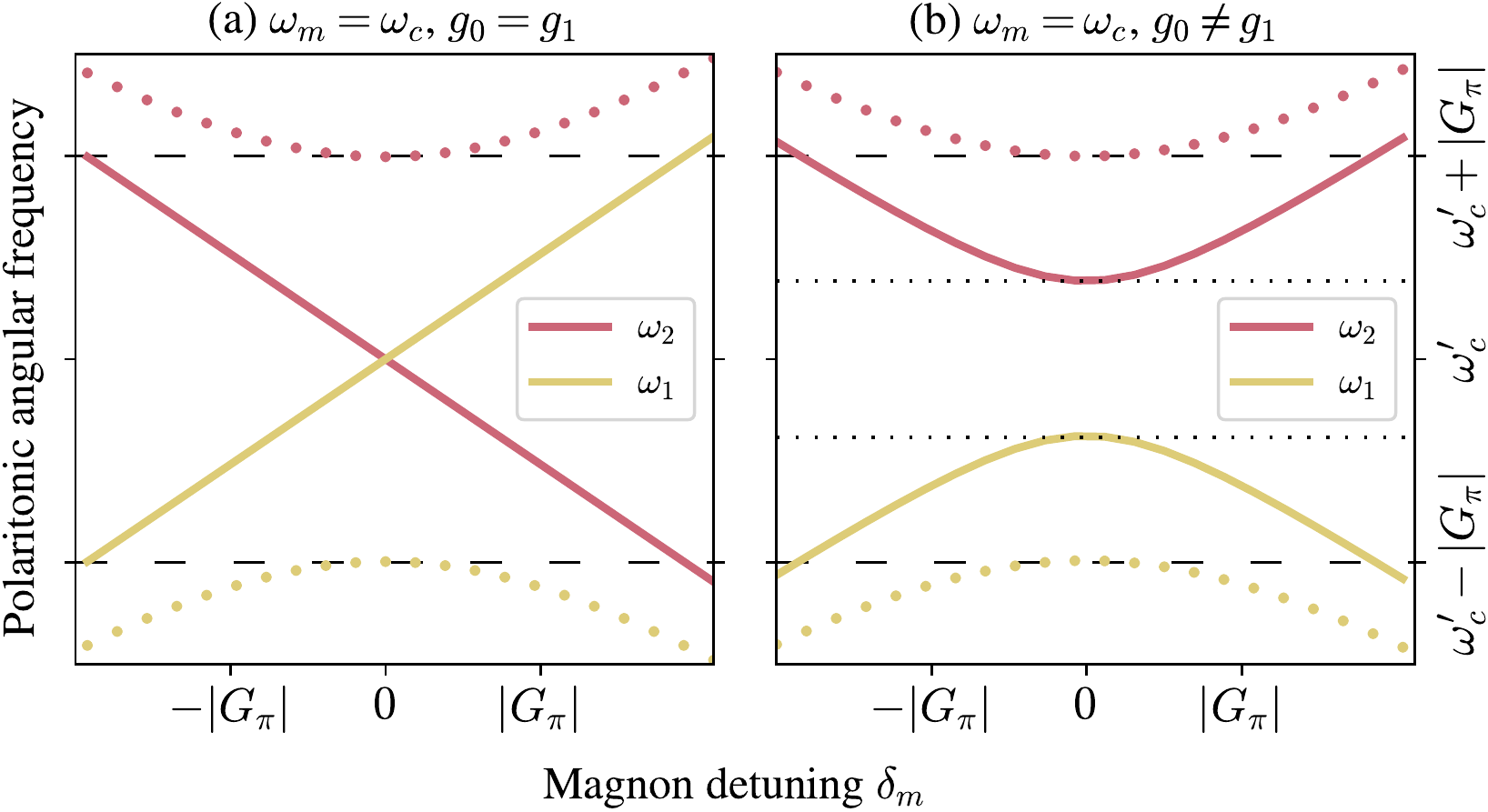}
		\caption{Numerical spectra of $H_\theta$ for $\theta=0$ (solid lines), and $\theta=\pi$ (dots) as a function of the magnon detuning $\delta_m$ at $\omega_m=\omega_c$. The dashed lines are at $\omega_c' \pm \abs{G_\pi}$ while the dotted lines are at $\omega_c' \pm \abs{G_0}$, with $G_\theta$ evaluated using \cref{eq:magnon-magnon-coupling} for $\delta_m=0$. We set $g_0=g_1 = 0.01 \omega_c$ in (a), $g_0 =0.01 \omega_{c, 0}$ and $g_1 = 0.01 \omega_{c, 1}$ in (b). The definition of $\omega_c'$ is per the main text. Other parameters and legend are as in \cref{fig:spectra}. }
		\label{fig:anticrossing}
		\phantomsubcaption\label{fig:anticrossing:middle-crossing}
		\phantomsubcaption\label{fig:anticrossing:middle-anticrossing}
	\end{figure}

	\paragraph{Cavity-mediated coupling.}
	To verify our analysis, we numerically plotted $\omega_1$ and $\omega_2$ by diagonalising \cref{eq:matrix-M-def} at $\omega_m=\omega_c$ in \cref{fig:anticrossing}. We observe the crossing of the solid lines (corresponding to $\theta=0$), while the dots ($\theta=\pi$) anti-cross. The situation changes if we set $g_0 \neq g_1$ as shown in \cref{fig:anticrossing}, and now we observe level repulsion for both $\theta=0$ and $\theta=\pi$. This behaviour is exactly the one we inferred in the dispersive regime. Note that our analysis of the dispersive regime is valid if $\abs{\frac{g_k}{\Delta_{kk'}}} \ll 1$. At $\omega_m=\omega_c$, this is equivalent to imposing $g_0,g_1 \ll \abs{\delta_c-\delta_m}$, which the parameters employed in \cref{fig:anticrossing} satisfy. Regardless of the values of $g_0$ and $g_1$, \cref{fig:anticrossing} shows that the frequency gap between the eigenmodes of \cref{eq:dispersive-hamiltonian-magnons} is minimal when $\omega'_{m,0} = \omega'_{m,1}$, i.e $\delta_m=0$. The location of this minimum gap is given by $\omega_c' = \omega_c + \frac{g_0^2-g_1^2}{\delta_c}$ due to the frequency shift of the magnon modes described by \cref{eq:wm-pulled}. 
	%The resulting frequency gap, characterising the strength of the indirect magnon-magnon coupling, is $2\abs{G_\theta}$. 
	
	Physically, \cref{eq:magnon-magnon-coupling} highlights an interference effect between the magnon-magnon couplings contributed by each cavity mode. This interference can be constructive for $\theta=\pi$, or destructive for $\theta=0$, and we always have $\abs{G_\pi}>\abs{G_0}$. In particular, if $g_0=g_1$, the cavity-mediated coupling between the magnons completely vanishes for $\theta=0$. We conclude that the case $\theta=0$ is always detrimental to indirectly couple two spatially distant magnons, while the case $\theta=\pi$ is benificial. This conclusion highlights the importance of the physical phase $\theta$ for multimode cavity-mediated coupling applications.

	\subsection{\label{subsec:dark-polaritons}Dark mode physics}
\begin{figure}[t]
		\centering
		\includegraphics[width=\linewidth]{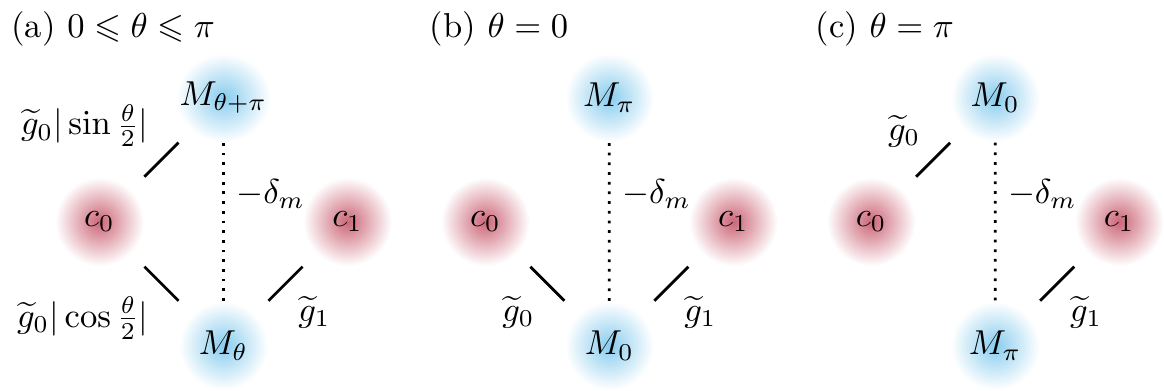}
		\caption{Schematic of the interactions present in $H_\theta$ in the rotated magnon basis (see \cref{eq:modelX-free:rotated,eq:modelX-interaction:rotated}) for (a) arbitrary $\theta$, (b) $\theta=0$, (c) $\theta=\pi$. The black lines are labelled by the magnitude of the couplings, and the dotted black line represents the possibility of vanishing $\delta_m$. Note that $M_{0}=M_{2\pi}$.}
		\label{fig:models}
		\phantomsubcaption\label{fig:modelX}
		\phantomsubcaption\label{fig:modelA}
		\phantomsubcaption\label{fig:modelB}
	\end{figure}

	\paragraph{Rotated magnon basis.} After having looked at the spectrum, we now turn our attention to the eigenmodes of $H_\theta$. The physics will be more easily understood by considering a rotated magnon basis ${M_\theta, M_{\theta+\pi}}$, defined by
	\begin{equation}
		\label{eq:rotated-magnon-basis}
		M_\theta = \frac{m_0 + e^{-i\theta} m_1}{\sqrt 2}.
	\end{equation}
	It is easy to prove that, since $m_0$ and $m_1$ commute with each other, so do $M_\theta$ and $M_{\theta+\pi}$. Note that $M_0=M_{2\pi}$ corresponds to co-rotating magnons (i.e. both magnons precess in-phase) while $M_\pi$ corresponds to counter-rotating magnons (precessing out-of-phase). In this basis, the free Hamiltonian \cref{eq:modelX-free-hamiltonian} now reads
	\begin{equation}
		\label{eq:modelX-free:rotated}
		\begin{aligned}
			H_\text{free} 
			&= 
			\hbar \omega_{c,0}c_0^\dagger c_0
			+ \hbar \omega_{c,1}c_1^\dagger c_1 \\
			&\quad
			+ \hbar \omega_m M_\theta^\dagger M_\theta 
			+ \hbar \omega_m M_{\theta+\pi}^\dagger M_{\theta+\pi}\\
			&\quad
			- \delta_m \qty(M_\theta M_{\theta+\pi}^\dagger + h.c).
		\end{aligned}
	\end{equation}
	while the interaction Hamiltonian \cref{eq:interaction-hamiltonian-general-single-phase} reads
	\begin{equation}
		\label{eq:modelX-interaction:rotated}
		\begin{aligned}
			H_{I,\theta} 
			&= 
			\hbar \widetilde{g}_0 e^{-i\theta/2} c_0 \qty(\cos \frac{\theta}{2}M_\theta - i\sin \frac{\theta}{2} M_{\theta+\pi})^\dagger\\
%			\hbar \widetilde{g}_0 e^{i\theta/2} \cos \frac{\theta}{2} c_0 M_\theta^\dagger\\
%			&\quad
%			+ \hbar \widetilde{g}_0 e^{i \frac{\theta-\pi}{2}} \sin \frac{\theta}{2} c_0 M_{\theta+\pi}^\dagger\\
			&\quad
			+ \hbar \widetilde{g}_1 c_1 M_\theta^\dagger + h.c,
		\end{aligned}
	\end{equation}
	where we introduced the notation $\widetilde{g}_k = \sqrt{2} g_k$. The different interaction pathways between the modes are illustrated in \cref{fig:modelX}. 
	
	Notably, if $\delta_m=0$, $H_\theta$ describes the interaction between four different modes, while for $\theta=0$ it separates into a 3-mode system $\qty{c_0,c_1, M_0}$ and a one-mode system $\qty{M_\pi}$, and for $\theta=\pi$ it separates into two two-mode systems $\qty{c_0,M_0}$ and $\qty{c_1, M_\pi}$. In particular, for $\delta_m=0$ and $\theta=0$ (\cref{fig:modelA}), we see that $M_\pi$ does not couple to any other mode.

	\paragraph{Dark mode definition.}
%	Interestingly, we see from \cref{eq:H0-rotated} and \cref{fig:polariton-magnon-phases:modelA-out} that for $\theta=0$ and $\delta_m=0$, $M_-$ does not couple to any other mode. 
	The presence of such an uncoupled eigenmode is reminiscent of dark mode physics.
	Dark (bright) states are states that weakly (strongly) couple with the readout mechanism. Due to their weak coupling, these states benefit from an increased lifetime over their bright counterpart, and hence are interesting candidates for storing information. Dark and bright modes are defined similarly, but refer to the associated polaritonic mode in continuous variable systems.
	
	The definition of a dark mode given above is dependent on the mechanism used to probe the system. Since here we are interested in photon/magnon hybridisation, we consider a detection based on a microwave transmission experiment mediated by photons. As a consequence, we expect that any eigenmodes coupling with cavity modes will indirectly couple to the readout medium, hence acquiring an experimental signature.
	
\begin{figure}[t]
		\centering
		\includegraphics[width=\linewidth]{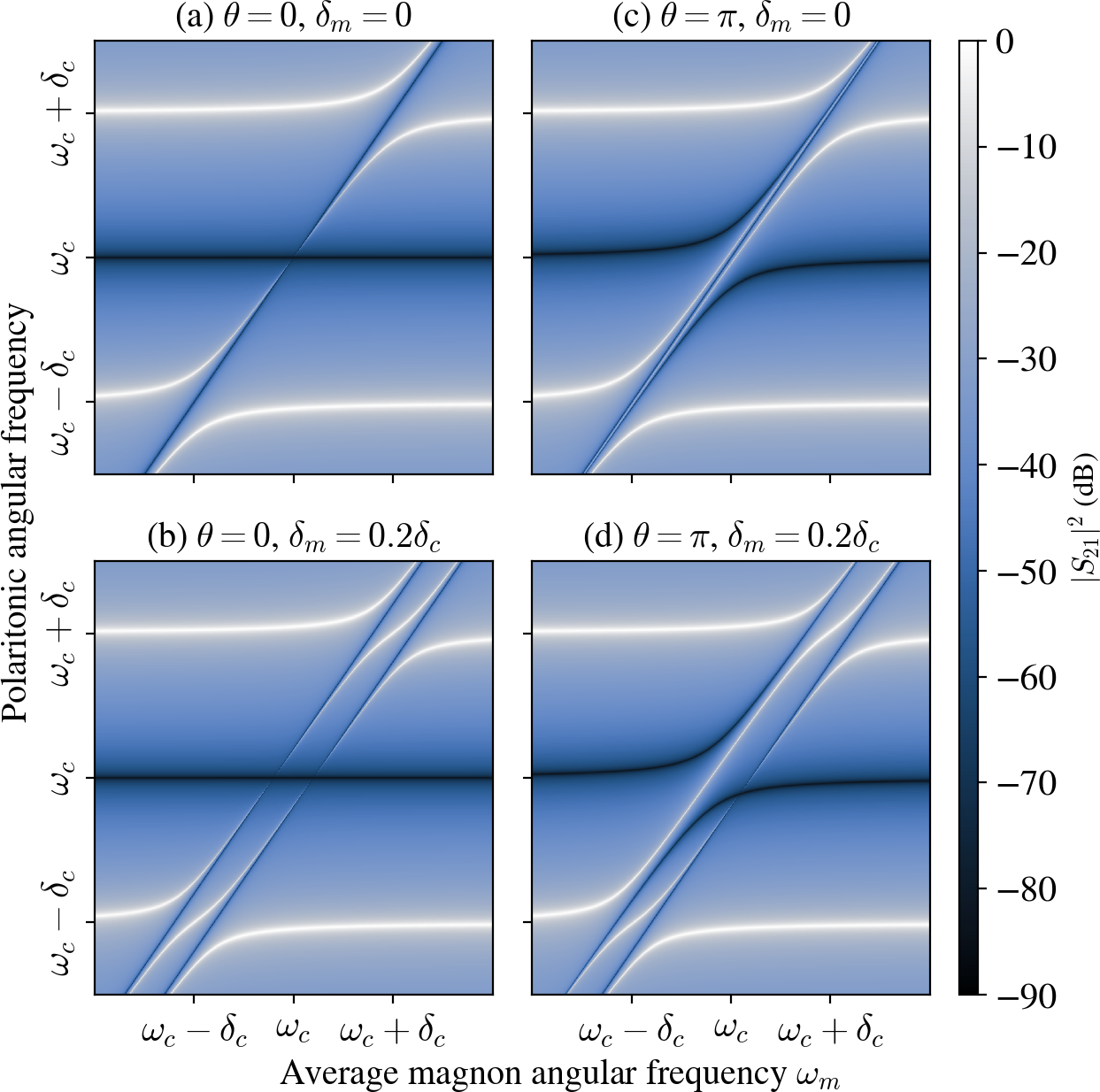}
		\caption{Numerical calculation of the microwave transmission through the system obtained using an input-output theory for $H_\theta$ at different magnon detuning $\delta_m$. See the main text for the parameters. In (a) we see that the spectral line associated to $\omega_1$ (see \cref{fig:spectra}) does not lead to a maximum of transmission, contrary to (b), (c) and (d). This shows that the eigenmode associated with $\omega_1$ is a dark mode.}
		\label{fig:input-output}
		\phantomsubcaption\label{fig:input-output:modelA}
		\phantomsubcaption\label{fig:input-output:modelA-detuned}
		\phantomsubcaption\label{fig:input-output:modelB}
		\phantomsubcaption\label{fig:input-output:modelB-detuned}
	\end{figure}

	The input-output formalism \cite{1985Gardiner} allows one to model this scenario. To illustrate our discussion of dark modes, we now consider the transmission through the cavity ($S_{21}$ parameter), experimentally obtained by attaching a vector network analyser to the two ports of a cavity. The derivation of the input-output theory is rather lengthy, so we refer the reader to \cref{app:input-output} for more details.
	We consider that each cavity mode $c_k$ couples to a photonic bath with coupling rate $\sqrt{\gamma/2\pi}$ and we set $\gamma= 5$ MHz. 
	We further introduce the intrinsic dissipation $\kappa = 1$ MHz of the cavity (due to its finite quality factor) and magnon modes (Gilbert damping) through $\widetilde{\omega}_{o,k}= \omega_{o,k} -i \kappa$. Other parameters are as in \cref{fig:spectra}. The numerical results are plotted in \cref{fig:input-output}.

	\paragraph{Dark mode physics.}
	Given the definition of a dark mode given above, we conclude that for $\theta=0$, $M_\pi$ is dark, since it is uncoupled to cavity modes. This is indeed what we observe in \cref{fig:input-output:modelA}: comparing to the spectrum in \cref{fig:spectra}, the polaritonic frequency $\omega_1/2\pi$ associated with $M_\pi$ does not give a maximum of transmission, unlike the other frequencies. This is in contrast with \cref{fig:input-output:modelB}, in which all the polaritonic frequencies lead to maxima of transmissions. Indeed, for $\theta=\pi$, $M_\pi$ couples to $c_1$. Hence, since all the eigenmodes for $\theta=\pi$ couple to photons, we conclude that strictly speaking, there are no dark modes for $\theta=\pi$. 
	
	These observations rely on both magnons modes having identical resonance frequencies, i.e. $\delta_m =0$. If this symmetry is broken by setting $\delta_m \neq 0$, a new interaction pathway between $M_\pi$ and $M_0$ opens (see \cref{fig:models}), regardless of the value of $\theta$. Since $M_0$ couples to the cavity modes, then $M_\pi$ will also couple to the cavity modes through $M_0$. As a consequence, we expect $M_\pi$ to acquire an experimental signature if $\delta_m \neq 0$. This theoretical prediction is supported numerically by \cref{fig:input-output:modelA-detuned,fig:input-output:modelB-detuned}. We note that this phenomenon of the illumination of a dark mode by symmetry breaking was already discussed by \cite{2022Nair} in the interaction picture.
	
	To conclude this section, we comment on the presence of two types of anti-resonances. The first type corresponds to dark diagonal lines which follow the FMR of the YIG spheres. They are the result of destructive interference between the FMR and the RF magnetic mode of the cavity, as explained by \cite{2016HarderHydeBaiMatchHu}. 
	The other type is due to destructive interference between the two cavity modes, and corresponds to the dark horizontal lines centred around $\omega_c = \frac{\omega_{c, 0}+\omega_{c, 1}}{2}$. These anti-resonances are uncoupled for $\theta=0$ (straight line for all $\omega_m$), while for $\theta=\pi$ an anti-crossing appears. This phenomenon can be explained by the fact that for $\theta=0$, both cavity modes couple to $M_0$ with equal coupling strengths, while the situation is different for $\theta=\pi$ (see \cref{fig:models}). Had we set $g_0 \neq g_1$, the anti-resonance due to the cavity modes would have also coupled near the FMR for $\theta=0$.

	\section{\label{sec:conclusion}Conclusion}
	To summarise our results, we showed that the interaction between a microwave photon and a magnon in a cavity is always associated with a coupling phase, which is different from that in dissipatively coupled systems. In most cases studied in the literature, this coupling phase can be omitted because of the simple topology of the interaction diagrams. Indeed, we traced the manifestation of the coupling phase to the presence of loops in the interaction diagrams (as illustrated by \cref{fig:phase-removal}), a phenomenon similar to a discrete Pancharatnam–Berry phase.
	To illustrate that the physics was dependent on the coupling phase, we considered a model composed of two cavity modes and two magnon modes, parametrised by the physical phase $\theta$. For different values of $\theta$, we found different behaviours with respect to two potential applications of cavity magnonics: one case is advantageous for cavity-mediated coupling but cannot be used for dark mode memories ($\theta=\pi$), and vice-versa for the other case ($
	\theta=0$). We note that while we showed results only for $\theta=0$ and $\theta=\pi$, our analytical derivations were still parametrised by a general $\theta$, and hence apply to intermediate cases. In fact, values of $\theta$ between 0 and $\pi$ continuously interpolate between the results presented in this paper (see the cavity-mediated coupling strength of \cref{eq:magnon-magnon-coupling} for example).
	
	One could question whether there are simpler models in which there is a physical phase. In theory, any system in which there are $n$ interactions between $n$ bosonic modes, with one being a microwave photon/magnon interaction, leads to a physical phase. One such example was considered by \citeauthor{2021Zhan}, where two ferromagnetic bilayers couple to a common cavity mode, but can also directly couple together thanks to the interlayer exchange interaction. This results in a triangular-shaped interaction diagram, so that in principle, a physical phase should parametrise their Hamiltonian. However, it is unclear how the impact of the physical phase could be experimentally verified in such a system. First, achieving different coupling phases for each ferromagnetic bilayers is a challenging experimental task, since the two layers need to be close to each other but be penetrated by the same cavity mode with a different angle (see \cref{eq:coupling-phase}). Furthermore, as noted by the authors of \citeauthor{2021Zhan}, the experimental verification of their analysis requires smaller magnon damping rates than currently reported in experiments. 
	
	Instead, the model introduced in \cref{sec:model} should be experimentally accessible using re-entrant cavity designs. Indeed, previous experimental results  demonstrated the tunability of the frequencies of the cavity modes, with values of the coupling strengths compatible with the RWA \cite{2020BourhillCastel,2016Carvalho,Tobar}. Placing two YIG spheres in a 3-post re-entrant cavity should allow to implement the two cases $\theta=0$ and $\theta=\pi$ discussed here, and will be the subject of a separate investigation.

	The engineering of cavities to achieve specific physical phases is an exciting research direction. The design of such cavities will consists in controlling the relative orientation of the magnetic modes traversing the magnetic samples, as described by \cref{eq:coupling-phase}. Finally, we note that we considered only a rather simple model in which a physical phase manifests. For instance, considering an additional cavity mode would lead to the appearance of two physical phases, for which the physics is yet to explored. More generally, the existence of discrete Pancharatnam–Berry phases in microwave cavity magnonics may find applications in quantum information processing.

	\begin{acknowledgments}
		We acknowledge financial support from Thales Australia and Thales Research and Technology. We thank Tyler Whittaker, Thomas Kong and Ross Monaghan for reading the manuscript and providing useful comments. The scientific colour map \textit{oslo} \cite{2021Crameri} is used in this study to prevent visual distortion of the data and exclusion of readers with colourvision deficiencies \cite{2020Crameri}. 
	\end{acknowledgments}

    \appendix
    \section{\label{app:derivation-form-factor}Photon/magnon coupling term}
    In this appendix, we show that the interaction term between a microwave cavity mode $c$ and a magnon mode $m$ can be expressed as
    \begin{equation}
        H_I
        =
        \hbar ge^{i\varphi} (c + c^\dagger) m^\dagger + h.c,
    \end{equation}
    with $g$ and $\varphi$ positive real numbers. This derivation is valid in the macrospin approximation \cite{2010SoykalPRB,2010SoykalPRL} and provided the number of magnon excitations is small compared to the number of spins.
    
    Our starting point is the supplementary material of ref. \cite{2019FlowerGoryachevBourhillTobar}. In particular, it is shown that in the macrospin approximation the interaction term is given by 
    \begin{equation}
    	\label{eq:interaction-Flower2019}
    	\begin{aligned}
    		H_I &= 
    		\hbar g^x \qty(c+c^\dagger)\qty(m + m^\dagger) + 
    		\hbar ig^y \qty(c+c^\dagger)\qty(m - m^\dagger) \\
    		&\quad+
    		\hbar g^z \qty(c+c^\dagger)m^\dagger m + \hbar \Omega^z \qty(c + c^\dagger)
    	\end{aligned}
    \end{equation}
	with 
	\begin{align}
		g^x
		&=
		-\frac{\gamma}{2 c V_m}
		\sqrt{\frac{\hbar \omega_c S}{\epsilon_0}}
		\frac{
			\int_{V_m} \vb{H} \cdot \vu{x} \dd[3]{r}
		}{
			\sqrt{\int_{V_c} \abs{\vb{H}}^2 \dd[3]{r}}
		},\\
		g^y
		&=
		\frac{\gamma}{2c V_m}
		\sqrt{\frac{\hbar \omega_c S}{\epsilon_0}}
		\frac{
			\int_{V_m} \vb{H} \cdot \vu{y} \dd[3]{r}
		}{
			\sqrt{\int_{V_c} \abs{\vb{H}}^2 \dd[3]{r}}
		},\\
		g^z
		&=
		\frac{\gamma}{cV_m}
		\sqrt{\frac{\hbar \omega_c}{2 \epsilon_0}}
		\frac{
			\int_{V_m} \vb{H} \cdot \vu{z} \dd[3]{r}
		}{
			\sqrt{\int_{V_c} \abs{\vb{H}}^2 \dd[3]{r}}
		},\\
		\Omega^z
		&=
		-\frac{\gamma S}{cV_m}
		\sqrt{\frac{\hbar \omega_c}{2 \epsilon_0}}
		\frac{
			\int_{V_m} \vb{H} \cdot \vu{z} \dd[3]{r}
		}{
			\sqrt{\int_{V_c} \abs{\vb{H}}^2 \dd[3]{r}}
		},
    \end{align}
	where $\gamma$ is the gyromagnetic ratio, $V_m$ the volume of the magnetic sample, $V_c$ the volume of the cavity, $\omega_c$ the resonance of the cavity, $S=\frac{\mu}{g_L \mu_B} N_S$ the total spin number of the macrospin operator, $\mu$ the magnetic moment of the magnetic sample, $g_L=2$ the Landé $g$-factor, $\mu_B$ the Bohr magneton, $N_s$ the number of spins in the magnetic sample, $c$ the speed of light in vacuum, $\epsilon_0$ the permittivity of vacuum, and $\vb{H}$ the magnetic mode of the cavity.
	
	Assuming that the magnetic mode has no $z$ dependence, $g^z = \Omega^z = 0$ and \cref{eq:interaction-Flower2019} can be written as $H_I = \hbar \qty(g^x -i g^y) \qty(c + c^\dagger) m^\dagger + h.c$ with
	\begin{align}
		g^x -i g^y
		&= 
		-\frac{\gamma}{2 c V_m}
		\sqrt{\frac{S \hbar \omega_c}{\epsilon_0}}
		\frac{
			\int_{V_m} \vb{H} \cdot \vu{x} \dd[3]{r} +
			i \int_{V_m} \vb{H} \cdot \vu{y} \dd[3]{r}
		}{\sqrt{\int_{V_c} \abs{\vb{H}}^2 \dd[3]{r}}} \\
		&= 
		\label{eq:app:tmp0}
		-\frac{\gamma}{2}
		\sqrt{\frac{\mu_0 S \hbar \omega_c}{V_m}}
		\frac{\widetilde{H}}{\sqrt{V_m \int_{V_c} \abs{\vb{H}}^2 \dd[3]{r}}} 
	\end{align}
	where we used $\mu_0 \epsilon_0 = 1/c^2$ and we recognise the quantity $\widetilde{H} = \int_{V_m} \vb{H} \cdot \vu{x} \dd[3]{r} + i \int_{V_m} \vb{H} \cdot \vu{y} \dd[3]{r}$ defined in \cref{eq:definition-H-tilde} of the main text. We introduced $\vb{H}$ as a cavity mode, which means that it is the solution of an eigenvalue problem for the electromagnetic field obeying the Maxwell's equations in the cavity. By definition, $-\vb{H}$ is also a solution to this eigenvalue problem, and hence \cref{eq:app:tmp0} is still valid after replacing $\vb{H} \mapsto - \vb{H}$. This removes the inconvenient minus sign \cref{eq:app:tmp0} so that we can now write
	\begin{align}
		H_I &=
		\frac{\hbar \gamma}{2}
		\sqrt{\frac{\mu_0 S \hbar \omega_c}{V_m}}
		\frac{\widetilde{H}}{\sqrt{V_m \int_{V_c} \abs{\vb{H}}^2 \dd[3]{r}}} 
		\qty(c + c^\dagger) m^\dagger + h.c\\
		&=
		\label{eq:app:tmp1}
		\frac{\hbar \gamma}{2}
		\sqrt{\frac{\mu_0 S \hbar \omega_c}{V_m}}
		\eta e^{i \theta}
		\qty(c + c^\dagger) m^\dagger + h.c
	\end{align}
	where we defined
	\begin{align}
		\eta 
		&= \frac{\abs{\widetilde{H}}}{\sqrt{V_m \int_{V_c} \abs{\vb{H}}^2 \dd[3]{r}}} \\
		&= \sqrt{
			\frac{
				\qty(\int_{V_{m}} \vb{H} \cdot \vu{x} \dd[3]{r})^2 + \qty(\int_{V_{m}} \vb{H} \cdot \vu{y} \dd[3]{r})^2
			}{V_{m} \int_{V_c} \abs{\vb{H}}^2 \dd[3]{r}}
		},\\
		\varphi &= \arg \widetilde{H}
	\end{align}
	as announced in the main text.
	
	To derive the expression of the coupling strength $g/2\pi$ as per the main text, we follow \cite{2020BourhillCastel} and write $S = \frac{\mu}{g_L \mu_B} N_s = \frac{\mu}{g_L \mu_B} n_s V_m$ with $n_s$ the spin density of the magnetic sample. \Cref{eq:app:tmp1} is then rewritten
	\begin{equation}
		H_I = \hbar g e^{i \varphi}
		\qty(c + c^\dagger) m^\dagger + h.c,
	\end{equation}
	with 
	\begin{equation}
		g/2\pi = \eta \sqrt{\omega_c} \frac{\gamma}{4\pi} \sqrt{\frac{\mu}{g_L \mu_B} \mu_0 \hbar n_s}.
	\end{equation}

	\section{\label{app:rotating-frame}Unitary transformations}	
	A time-dependent unitary transformation $U$ maps a Hamiltonian $H$ to $H'$ such that
	\begin{equation}
		\label{eq:time-dependent-untary-transformation-law}
		H \mapsto H' = UHU^\dagger + i\hbar \dot{U} U^\dagger.
	\end{equation}
	This corresponds to the so-called rotating frame transformations. For a time-independent unitary transformation, $H' = UHU^\dagger$, we simply refer to it as a change of frame.
	The Baker–Campbell–Hausdorff formula is useful for computing the transformed Hamiltonian, it reads
	\begin{equation}
		\label{eq:Baker-Campbell-Hausdorff}
		e^{X}Ye^{-X} = \sum_{n=0}^\infty \frac{\qty[X^{(n)}, Y]}{n!}
	\end{equation}
	where $X$ and $Y$ are operators, and we note $\qty[\qty(X)^{(n)}, Y]$ the $n$-th iterated commutator defined as
	\begin{gather}
		\qty[X^{(0)}, Y] \equiv Y, \quad
		\qty[X^{(1)}, Y] = \qty[X, Y] = XY - YX\\
		\qty[X^{(2)}, Y] = \qty[X, \qty[X, Y]], \quad
		\qty[X^{(3)}, Y] = \qty[X, \qty[X^{(2)}, Y]]
	\end{gather}
	and so on.
	
	For an annihilation operator $a$ following bosonic commutation relation $\qty[a, a^\dagger]=1$ and $\varphi \in \mathbb{R}$, the unitary transformation $U = e^{i \varphi a^\dagger a}$ transforms $a$ and $a^\dagger$ as
	\begin{align}
		UaU^\dagger &= e^{i\varphi a^\dagger a}ae^{-i\varphi a^\dagger a} = a e^{-i \varphi},\\
		Ua^\dagger U^\dagger &=  a^\dagger e^{i \varphi}.
	\end{align}
	
	The associated number operator $a^\dagger a$ is unaffected by the transformation, since
	\begin{align}
		Ua^\dagger aU^\dagger &= Ua^\dagger U^\dagger U aU^\dagger = a^\dagger e^{i \varphi} a e^{-i \varphi} \\
		&= a^\dagger a
	\end{align}
	where $U^\dagger U = I$ since $U$ is unitary. 
	
	Additionally, for any operator $O$ that commutes with $a^\dagger a$, i.e $\qty[a^\dagger a, O]=0$, then $O$ is unaffected by the transformation, i.e $UOU^\dagger = O$, due to the vanishing of the iterated commutators in the Baker–Campbell–Hausdorff formula (\cref{eq:Baker-Campbell-Hausdorff}).

	\section{\label{app:relevance-phase}Relevance of the phase factor}
	\subsection{Removal of the coupling phase for a magnon/photon system}
\begin{center}
		\begin{figure}[h]
			\includegraphics[width=\linewidth]{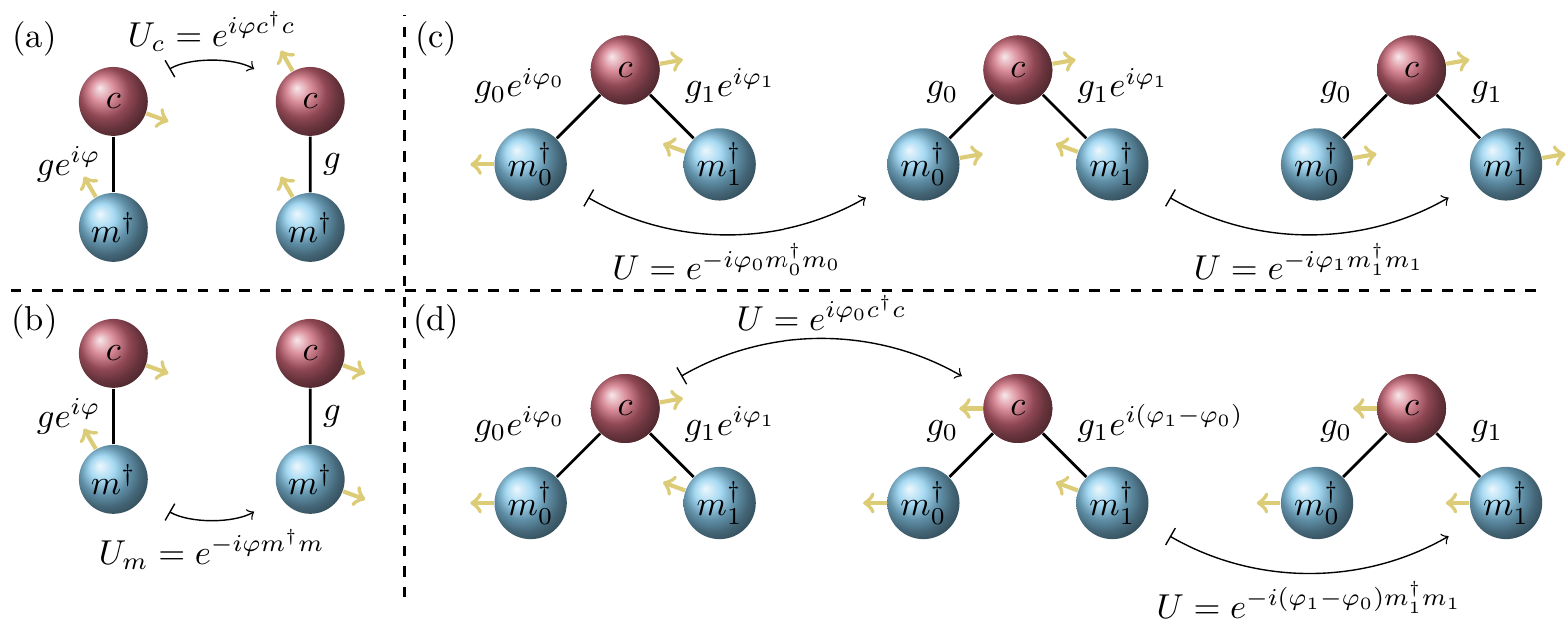}
			\caption{Example diagrams showing coupling phase removal by unitary transformations. The conventions used are that of \cref{fig:physical-phase}. (a) Rotation of the cavity mode $c$ using $U_c=e^{i \varphi c^\dagger c}$. The magnon mode $m$ is unaffected. (b) Rotation of the magnon mode $m$ using $U_m=e^{-i \varphi m^\dagger m}$. (c) Rotation of both magnon modes using $U = e^{-i\varphi_j m_j^\dagger m_j}$ for $j \in \qty{0, 1}$. (d) Rotation of the cavity mode $c$ using $U = e^{i\varphi_0 c^\dagger c}$ and the magnon mode $m_1$ using $U=e^{-i\qty(\varphi_1-\varphi_0) m_1^\dagger m_1}$.}
			\label{fig:phase-removal}
			\phantomsubcaption\label{fig:cm:rotate-c}
			\phantomsubcaption\label{fig:cm:rotate-m}
			\phantomsubcaption\label{fig:cmm:rotate-m0-m1}
			\phantomsubcaption\label{fig:cmm:rotate-c-m1}
		\end{figure}
	\end{center}

	In the simple case of a single cavity mode coupling to a single magnon mode, the Hamiltonian is $H = H_\text{free} + H_I$ defined by \cref{eq:free-hamiltonian-two-modes,eq:interaction-complex-simple}. 
	In light of \cref{fig:cm:rotate-c}, we can \onecolumngrid
	
	\twocolumngrid\noindent	choose to rotate the cavity mode by $\varphi$ using $U_c = e^{i\varphi c^\dagger c}$. Formally, $H_\text{free} \mapsto H_\text{free}' = H_\text{free}$ is unchanged while
	\begin{align}
		H_I \mapsto H_I' 
		&=  e^{i\varphi c^\dagger c} \qty(\hbar ge^{i\varphi} c m^\dagger + h.c) e^{-i\varphi c^\dagger c}\\
		&= \hbar ge^{i\varphi} \qty(e^{i\theta c^\dagger c}ce^{-i\varphi c^\dagger c}) 
		\qty(e^{i\theta c^\dagger c} m^\dagger e^{-i\varphi c^\dagger c}) + h.c\\
		&= \hbar ge^{i\varphi} \qty(c e^{-i\varphi}) m^\dagger + h.c\\
		&= \hbar gc m^\dagger + h.c
	\end{align}
	
	hence removing the coupling phase $\varphi$. Pictorially, this is illustrated on the right of \cref{fig:cm:rotate-c}, where we see that the yellow arrow of mode $c$ has changed orientation, and is now pointing in the same direction as that of mode $m$.
	Alternatively, as illustrated in \cref{fig:cm:rotate-m}, we can rotate the magnon mode instead, using $U_m = e^{-i\varphi m^\dagger m}$, which yields the same result, albeit with a different final orientation of the yellow arrows.
	
	The fact that $\varphi$ is absent after a change of frame implies that it cannot be physical. Similarly, note that the two extremal cases $\varphi=0$ and $\varphi=\pi$ differ only by the sign of the coupling strength, and hence we deduce that its polarity does not affect the physics.

	\subsection{Removal of the coupling phase for two magnons coupling to a common cavity mode}
	%\paragraph{System definition and relevance.}
	We now assume that an additional magnon mode couples to the cavity mode, as considered in many works \cite{2015Zhang,2016Lambert,2018Rameshti,2019Grigoryan,2021Zhan,2022Nair,2019LiZhu,2019ZhangScullyAgarwal,2020Nair}. 
	We also note that since all modes are bosonic, this scenario is equivalent to one where a single magnon mode couples to two cavity modes, as considered e.g by \cite{2016HarderHydeBaiMatchHu}. 
	The Hamiltonian is $H = H_\text{free} + H_I$ with
	\begin{align}
		H_\text{free} &=
		\hbar \omega_{c} c^\dagger c
		+ \hbar \omega_{m,0} m_0^\dagger m_0
		+ \hbar \omega_{m,1} m_1^\dagger m_1,\\
		H_I &=
		\hbar g_0e^{i\varphi_0} c m_0^\dagger + \hbar g_1e^{i\varphi_1} c m_1^\dagger + h.c,
	\end{align}
	and is illustrated in \cref{fig:cmm:rotate-m0-m1,fig:cmm:rotate-c-m1}.
	
	%\paragraph{Removal of the phase.}
	One simple possibility to remove the coupling phases $\varphi_i$ is to successively rotate $m_i$ using $U_{m_i} = e^{-i\varphi_i m_i^\dagger m_i}$ for $i \in \qty{0, 1}$ (\cref{fig:cmm:rotate-m0-m1}). 
	While this is the simplest method, we could also rotate the cavity mode $c$ (\cref{fig:cmm:rotate-c-m1}), for instance to remove $\varphi_0$ using $U_c = e^{i\varphi_0 c^\dagger c}$, but then the coupling to the second magnon is also affected and reads $\hbar g_1e^{i\qty(\varphi_1-\varphi_0)} c m_1^\dagger + h.c$. 
	Applying $U_{m_1} = e^{-i\qty(\varphi_1-\varphi_0) m_1^\dagger m_1}$ removes the remaining phase.
	
	%\paragraph{Interpretation of the physics after rotation.}
	Formally, it appears that the coupling phases are again unimportant for the resulting physics. While the physics is unchanged, we would like to mention that the \emph{interpretation} of the physics is: the rotated frame is not the laboratory frame. As an example, \cite{2022Nair} studied the specific case of $\varphi_0=\varphi_1=0$ and $g_0=g_1=g$, and found that $M_- = \frac{m_0-m_1}{\sqrt 2}$ was the only dark mode of the system. If we were to consider $g_0=-g_1=g$ instead (obtained by applying $U=e^{-i\pi m_1^\dagger m_1}$ for instance), we would find that the dark mode is transformed to $M_+ = \frac{m_0+m_1}{\sqrt 2}$ in this frame. Still, the spectrum and other physical predictions of \cite{2022Nair} would all hold.

	\subsection{Two cavity modes and two magnon modes}
	We note that the removal of the coupling phases in the previous examples was facilitated by the fact that the interaction diagrams had open ends: the rotation of the modes at the ends of the interaction chain would at most affect one edge.  Having a closed loop, however, removes this freedom, resulting in an observable physical phase $\theta$. We believe the hybrid system introduced in \cref{sec:model} is the simplest system with an interaction loop involving a magnon/photon interaction, which could be experimentally accessible.

	\section{\label{app:schrieffer-wolff}Effective Hamiltonian in the dispersive regime}
	For the purpose of this section, we adopt different notations for convenience. The model studied in the main 
	text can be written as $H_\theta = H_\text{free} + H_I$
	\begin{align}
		\label{eq:schrieffer-wolff:free}
		H_\text{free} &=
		\sum_{k=0}^1 \hbar \omega_{c,k} c_k^\dagger c_k + \hbar \omega_{m,k} m_k^\dagger m_k\\
		\label{eq:schrieffer-wolff:int}
		H_I &= \sum_{k,k'=0}^1 \hbar g_{kk'} c_k m_{k'}^\dagger + h.c
	\end{align}
	
	with $g_{0k'} = g_0$, $g_{10}=g_1$, and $g_{11} = g_1 e^{i\theta}$. We will hence assume that $g_{kk'} \in \mathbb{C}$ in this section, even though we will mostly be interested in the cases $\theta =0$ or $\pi$, for which the $g_{kk'}$ become reals. We also introduce the detuning $\Delta_{k k'} = \omega_{m, k'} - \omega_{c,k}$ between the cavity mode $c_k$ and the magnon mode $m_{k'}$.	In the rest of this section, we assume that the magnons are significantly detuned from the both cavity mode such that $\abs{\lambda_{k k'}} \ll 1$ with $\lambda_{kk'} = \frac{g_{k k'}}{\Delta_{k k'}}$ for $k,k' \in \qty{0,1}$. 
	
	To find the effective Hamiltonian in this regime, we employ a Schrieffer-Wolff transformation \cite{1966Schrieffer} to treat $H_I$ as a perturbation of $H_\text{free}$. Using the unitary transformation $U = e^{\Lambda}$ with
	\begin{equation}
		\Lambda = \sum_{k,k'} \qty(\lambda_{k k'} c_k m_{k'}^\dagger - \lambda_{k k'}^* c^\dagger_k m_{k'}),
	\end{equation}
	we have $H_I + \qty[\Lambda, H_\text{free}] = 0$ and the Baker–Campbell–Hausdorff formula (see \cref{eq:Baker-Campbell-Hausdorff}) gives $H'_\theta = e^{\Lambda} H_\theta e^{-\Lambda} = H_\text{free} + \frac{1}{2} \qty[\Lambda, H_I]$ to first order in $\abs{\lambda}$.
	
	After using commutator identities, we find the commutator 
	\begin{equation}
		\label{eq:schrieffer-wolff:commutator}
		\begin{aligned}
			[\Lambda, H_I]
			&=
			2\hbar \sum_{k} \qty[
				\qty(\sum_l \frac{\abs{g_{lk}}^2}{\Delta_{lk}}) m_{k}^\dagger m_{k}
				-
				\qty(\sum_l \frac{\abs{g_{kl}}^2}{\Delta_{kl}}) c_k^\dagger c_{k}
			]\\
			&\quad
			+\hbar \sum_{k\neq k'}\qty[
				\qty(\sum_l \lambda_{l k'} g_{l k}^*) m_{k} m_{k'}^\dagger
				+ h.c
			]\\
			&\quad
			-\hbar \sum_{k\neq k'}\qty[
				\qty(\sum_l \lambda_{k l} g_{k' l}^*) c_k c_{k'}^\dagger
				+ h.c
			] + \text{cst}.
		\end{aligned}
	\end{equation}
	The constant originates from using the bosonic commutation relations for the operators $c_k,m_k$.
	Hence, after applying $U = e^{\Lambda}$, we obtain the effective Hamiltonian
	\begin{equation}
		\label{eq:disperive-hamiltonian}
		\begin{aligned}
			H'_\theta 
			&=
			\sum_k \hbar \omega_{c, k}' c_k^\dagger c_k + \hbar \omega_{m, k}' m_k^\dagger m_k\\
			&\quad
			+ \sum_{k \neq k'} \qty(\kappa_{kk'} c_k c_{k'}^\dagger + h.c)
			+ \sum_{k \neq k'} \qty(G_{kk'} m_k m_{k'}^\dagger + h.c)
		\end{aligned}
	\end{equation}
	with shifted resonances
	\begin{align}
		\omega_{c,k}' &= \omega_{c, k} - \sum_{k'} \frac{\abs{g_{kk'}}^2}{\Delta_{kk'},}\\
		\omega_{m,k}' &= \omega_{m, k} + \sum_{k'} \frac{\abs{g_{k'k}}^2}{\Delta_{k'k}},
	\end{align}
	photon/photon couplings
	\begin{equation}
		\kappa_{kk'} = -\sum_l \frac{g_{k l} g_{k' l}^*}{2\Delta_{kl}},
	\end{equation}
	and magnon/magnon couplings
	\begin{equation}
		G_{kk'} = \sum_l \frac{g_{l k'} g_{l k}^*}{2\Delta_{lk'}}.
	\end{equation}

	The effective Hamiltonian highlights the decoupling of the photonic and matter degrees of freedom in the dispersive regime: light and matter do not interact anymore. However, we obtained a photon/photon coupling mediated by virtual magnons, nad magnon/magnon couplings mediated by virtual photons.
	
	Concentrating on the magnon modes, and setting $\omega_m = \omega_c$, we obtain the effective Hamiltonian for the magnons
	\begin{equation}
		H_m' = \sum_k \hbar \omega_{m, k}' m_k^\dagger m_k + \hbar \qty(G_\theta m_0 m_1^\dagger + h.c)
	\end{equation}
	with $G_\theta = G_{01} + G_{10}^*$
%	\begin{equation}
%		G =	\qty(g_0^2-e^{i\theta}g_1^2) \frac{\delta_c}{\delta_c^2-\delta_m^2},
%	\end{equation}
	and shifted magnon frequencies
	\begin{align}
		\omega_{m,0}'
		&=
		\omega_c + \frac{\qty(g_0^2 + g_1^2) \delta_m + \qty(g_0^2-g_1^2) \delta_c}{\delta_c^2 - \delta_m^2},\\
		\omega_{m,1}'
		&=
		\omega_c - \frac{\qty(g_0^2+g_1^2) \delta_m - \qty(g_0^2 - g_1^2) \delta_c}{\delta_c^2 - \delta_m^2}.
	\end{align}

    \section{\label{app:input-output}Details of the input-output theory}
    In the main text, we consider the transmission of the cavity by microwave photons.  In this appendix, we derive an analytical expression of the transmission of the cavity using the input-output formalism \cite{1985Gardiner}. We note that similar results can be obtained employing the loop theory introduced in \cite{2020Yuan}. In the following, we assume that we have replaced $\omega_{o, k} \mapsto \omega_{o,k} - i \kappa$ for $o_k \in \qty{c,m}, k \in \qty{0, 1}$ with $\kappa$ the intrinsic dissipation rate.
    
    \paragraph{Bath modelling.}
	We model the environment with a bosonic bath Hamiltonian $H_\text{bath}$ with operators $a_\omega$ and $b_\omega$ for the left and right side of the cavity. It reads
	\begin{equation}
		H_\text{bath} = \int_{-\infty}^\infty \dd{\omega} \qty(
		\hbar \omega a_\omega^\dagger a_\omega +
		\hbar \omega b_\omega^\dagger b_\omega
		).
	\end{equation}
	The interaction between the bath and the cavity is
	\begin{equation}
		H_\text{sys-bath} = i \hbar \int_{-\infty}^\infty \dd{\omega} 
		\qty(\sum_{k=0}^1 c_k \qty(\kappa_{a,k} a^\dagger_\omega + \kappa_{b,k} b^\dagger_\omega) + h.c),
	\end{equation}
	where we define $\sqrt{\gamma_{o,k}/2\pi} = \kappa_{o,k}$ for the bath operators $o \in \qty{a, b}$.
	The Heisenberg equation of motion for the bath operators $o \in \qty{a, b}$ is
	\begin{equation}
		\dot{o}_\omega
		= -\frac{i}{\hbar}\qty[o_\omega, H_\text{bath} + H_\text{sys-bath}]
		=
		-i\omega o_\omega + \sum_k \kappa_{o,k} c_k,
	\end{equation}	
	with the formal solutions
	\begin{equation}
		\label{eq:app:solution-t0}
		\begin{aligned}
			o_\omega(t) &= o_\omega(t_0)e^{-i\omega(t-t_0)} \\
			&\quad+ \sum_k \kappa_{o,k} \int_{t_0}^t \dd{t'} c_k(t') e^{-i\omega(t-t')}
		\end{aligned}
	\end{equation}
	for $t_0 < t$ and
	\begin{equation}
		\begin{aligned}
			o_\omega(t) &= o_\omega(t_1)e^{-i\omega(t-t_1)} \\
			&\quad - \sum_k \kappa_{o,k} \int_{t}^{t_1} \dd{t'} c_k(t') e^{-i\omega(t-t')}
		\end{aligned}
	\end{equation}
	for $t < t_1$.
%	\onecolumngrid
%	\begin{align}
%		\label{eq:app:solution-t0}
%		o_\omega(t) &= o_\omega(t_0)e^{-i\omega(t-t_0)} + \sum_k \kappa_{o,k} \int_{t_0}^t \dd{t'} c_k(t') e^{-i\omega(t-t')}, \quad t_0 < t\\
%		o_\omega(t) &= o_\omega(t_1)e^{-i\omega(t-t_1)} - \sum_k \kappa_{o,k} \int_{t}^{t_1} \dd{t'} c_k(t') e^{-i\omega(t-t')}, \quad t<t_1.
%	\end{align}

%	\twocolumngrid
	\paragraph{System for the cavity modes in frequency space.}
	We now define the input fields
	\begin{equation}
		o^\text{in}(t) 
		= \frac{1}{\sqrt{2\pi}} \int \dd{\omega} e^{-i\omega(t-t_0)} o_\omega(t_0)
	\end{equation}
	and using the solution of \cref{eq:app:solution-t0} for $t_0<t$ we have
	\begin{equation}
		\label{eq:input-output:tmp-o-in}
		\kappa_{o,k}^* \int_\mathbb{R} \dd{\omega} o_\omega(t)
		=
		\sqrt{\gamma_{o,k}^*}  o^\text{in}(t)
		+ \sum_{k'} \frac{\sqrt{\gamma_{o,k}^* \gamma_{o,k'}}}{2} c_{k'}(t)
	\end{equation}
	We adopt the notations of \cref{app:schrieffer-wolff} and write the interaction term $H_I = \sum_{k,k'=0}^1 g_{kk'} c_k m_{k'}^\dagger + h.c$. 
	The Heisenberg equation for the cavity modes at $t_0 < t$ reads
	\begin{align}
		\dot{c}_k
		&= -\frac{i}{\hbar}\qty[c_k, H_\text{sys} + H_\text{sys-bath}]\\
		&\begin{gathered}
			= -i\omega_{c,k} c_k 
			- i\sum_{k'} g_{kk'} m_{k'} 
			- \sum_{o \in \qty{a, b}} \sqrt{\gamma_{o,k}^*} o^\text{in}(t)\\
			- \sum_{o \in \qty{a, b}, k'} \frac{\sqrt{\gamma_{o,k}^* \gamma_{o,k'}}}{2} c_{k'}(t)
		\end{gathered}
	\end{align}	
	while that for the magnon modes is
	\begin{align}
		\dot{m}_k
		&= -\frac{i}{\hbar}\qty[m_k, H_\text{sys}]
		=
		-i\omega_{m,k} m_k 
		- i\sum_{k'} g_{k'k} c_{k'}.
	\end{align}	
	In frequency space, 
	\begin{equation}
		-i \omega \widetilde{m}_k
		=
		-i\omega_{m,k} \widetilde{m}_k 
		- i\sum_{k'} g_{k'k} \widetilde{c}_{k'}
	\end{equation}	
	which gives
	\begin{equation}
		\widetilde{m}_k = \frac{\sum_{k'} g_{k'k} \tilde{c}_{k'}}{\omega - \omega_{m,k}}.
	\end{equation}
	
	At this stage, it is convenient to introduce the detunings
	\begin{equation}
		\Delta_{o,k} = \omega - \omega_{o,k}, \quad o \in \qty{c,m}, \, k \in \qty{0,1}.
	\end{equation}
	The cavity modes in Fourier space then read
	\begin{equation}
%		-i \omega \tilde{c}_k 
%		=
%		-i\omega_{c,k} \tilde{c}_k 
%		- i\sum_{k'} g_{kk'} \tilde{m}_{k'} 
%		- \sqrt{\gamma_{o,k}^*} \tilde{o}^\text{in}
%		- \sum_{k'} \frac{\sqrt{\gamma_{o,k}^* \gamma_{o,k'}}}{2} \tilde{c}_k\\
%		\Delta_{c,k} \tilde{c}_{k'}
%		=
%		\sum_{k'} g_{kk'} \tilde{m}_{k'} 
%		- i \sqrt{\gamma_{o,k}^*} \tilde{o}^\text{in}
%		- i\sum_{k'} \frac{\sqrt{\gamma_{o,k}^* \gamma_{o,k'}}}{2} \tilde{c}_{k'}\\
%		\Delta_{c,k} \tilde{c}_k +
%		i\sum_{k'} \frac{\sqrt{\gamma_{o,k}^* \gamma_{o,k'}}}{2} \tilde{c}_{k'}
%		=
%		\sum_{k',l} \frac{g_{kk'} g_{lk'}}{\Delta_{m,k'}}\tilde{c}_{l}
%		- i \sqrt{\gamma_{o,k}^*} \tilde{o}^\text{in}\\
	\begin{gathered}
		\Delta_{c,k} \tilde{c}_k +
		\sum_{o \in \qty{a,b},k'} \qty(
		i \frac{\sqrt{\gamma_{o,k}^* \gamma_{o,k'}}}{2} 
		- \sum_{l} \frac{g_{kl} g_{k'l}}{\Delta_{m,l}}
		)\tilde{c}_{k'}\\
		+ \sum_{o \in \qty{a,b}} i \sqrt{\gamma_{o,k}^*} \tilde{o}^\text{in} = 0.
	\end{gathered}
	\end{equation}
	\emph{Separating the terms $k=k'$ and $k \neq k'$}, we obtain the system
	\begin{align}
		A_0 \tilde{c}_0 + B_{01} \tilde{c}_{1} + \sum_{o \in \qty{a,b}} O_{o,0} \tilde{o}^\text{in} &= 0\\
		B_{10} \tilde{c}_{0} + A_1 \tilde{c}_1 + \sum_{o \in \qty{a,b}} O_{o,1} \tilde{o}^\text{in} &= 0
	\end{align}
	with the solutions
	\begin{align}
		\label{eq:solution-c0}
		\tilde{c}_0 &=
		\sum_{o \in \qty{a,b}} \frac{O_{o,0} A_1 - O_{o,1} B_{01}}{B_{10}B_{01} - A_0 A_1} \tilde{o}^\text{in}\\
		\label{eq:solution-c1}
		\tilde{c}_1 &=
		\sum_{o \in \qty{a,b}} -\frac{O_{o,0} B_{10} - O_{o,1} A_0}{B_{10}B_{01} - A_0 A_1} \tilde{o}^\text{in}
	\end{align}
%	we can write
%	\begin{equation}
%		\qty(
%		\Delta_{c,k} 
%		+ i\frac{\abs{\gamma_{o,k}}}{2}
%		- \frac{g_{k0}g_{k0}}{\Delta_{m,0}} 
%		- \frac{g_{k1}g_{k1}}{\Delta_{m,1}} 
%		)\tilde{c}_k +
%		\qty(
%		i\frac{\sqrt{\gamma_{o,k}^* \gamma_{o,k'}}}{2}
%		- \frac{g_{k0}g_{k'0}}{\Delta_{m,0}} 
%		- \frac{g_{k1}g_{k'1}}{\Delta_{m,1}} 
%		)\tilde{c}_{k'} +
%		i \sqrt{\gamma_{o,k}^*} \tilde{o}^\text{in} = 0
%	\end{equation}
%	and hence
%	\begin{equation}
%		A_k \tilde{c}_k + B_{kk'} \tilde{c}_{k'} + O_k \tilde{o}^\text{in} = 0
%	\end{equation}
	and
	\begin{align}
		A_k 
		&= 
		\qty(
		\Delta_{c,k}
		+i \sum_{o \in \qty{a,b}} \frac{\abs{\gamma_{o,k}}}{2}
		) \Delta_{m,0}\Delta_{m,1} \nonumber\\
		&\quad
		- g_{k0}^2\Delta_{m,1}
		- g_{k1}^2\Delta_{m,0},\\
		B_{kk'}
		&=
		i \qty(\sum_{o \in \qty{a,b}} \frac{\sqrt{\gamma_{o,k}^* \gamma_{o,k'}}}{2})\Delta_{m,0}\Delta_{m,1} \nonumber\\
		&\quad
		- g_{k0}g_{k'0} \Delta_{m,1}
		- g_{k1}g_{k'1} \Delta_{m,0},\\
		O_{o,k} &= i \sum_{o \in \qty{a,b}} \sqrt{\gamma_{o,k}^*} \Delta_{m,0}\Delta_{m,1}.
	\end{align}

	\paragraph{Input-output relations.}
	We now define the output field
	\begin{equation}
		o^\text{out}(t) 
		= \frac{1}{\sqrt{2\pi}} \int \dd{\omega} e^{-i\omega(t-t_1)} o_\omega(t_1),
	\end{equation}
	and using the solution at $t<t_1$ we have
	\begin{align}
		\kappa_{o,k}^* \int_\mathbb{R} \dd{\omega} o_\omega(t) 
		%		&=
		%		\sqrt{\gamma_{o,k}^*} o^\text{in}(t)
		%		+ \sum_{k'} \frac{\sqrt{\gamma_{o,k}^* \gamma_{o,k'}}}{2} c_{k'}(t)\\
		&=
		\sqrt{\gamma_{o,k}^*} o^\text{out}(t)
		- \sum_{k'} \frac{\sqrt{\gamma_{o,k}^* \gamma_{o,k'}}}{2} c_{k'}(t)
	\end{align}
	which combined with \cref{eq:input-output:tmp-o-in} leads to input-output relations
%	\begin{align}
%		\sqrt{\gamma_{o,k}^*} o^\text{out}(t) - \sqrt{\gamma_{o,k}^*} o^\text{in}(t)
%		=
%		2\sum_{k'} \frac{\sqrt{\gamma_{o,k}^* \gamma_{o,k'}}}{2} c_{k'}(t).
%	\end{align}
%	We then obtain the input-output relations
	\begin{equation}
		\tilde{o}^\text{out} - \tilde{o}^\text{in} = \sum_{k} \sqrt{\gamma_{o,k}} \tilde{c}_k
	\end{equation}
	valid for each bath operator $o \in \qty{a,b}$.

	\paragraph{Transmission.}
	The transmission through the system can be defined as 
	\begin{equation}
		S_{21} = \left.\frac{\expval{\tilde{b}^\text{out}}}{\expval{\tilde{a}^\text{in}}}\right|_{\expval{\tilde{b}^\text{in}}=0}
		= \frac{\sum_k \sqrt{\gamma_{b,k}}\expval{ \tilde{c}_k}}{\expval{\tilde{a}^\text{in}}}
	\end{equation}
	where we used the input-output relations. Using \cref{eq:solution-c0,eq:solution-c1}, we obtain
	%$S_{21} = \left.\frac{\expval{\tilde{b}^\text{out}}}{\expval{\tilde{a}^\text{in}}}\right|_{\expval{\tilde{b}^\text{in}}=0}$. 
	%Using the input-output relations, and replacing the cavity modes $\tilde{c}_k$, the transmission reads
	\begin{equation}
		S_{21} 
		= 
		\sqrt{\gamma_{b,0}} \frac{O_{a,0} A_1 - O_{a,1} B_{01}}{B_{10}B_{01} - A_0 A_1}
		-\sqrt{\gamma_{b,1}} \frac{O_{a,0} B_{10} - O_{a,1} A_0}{B_{10}B_{01} - A_0 A_1}.
	\end{equation}
	
	\bibliography{coupling-phase}
	
\end{document}